\documentclass[showkeys,superscriptaddress,floatfix,prd,10pt,aps]{revtex4-2}
\usepackage{graphicx,epstopdf}
\usepackage[caption=false]{subfig}
\usepackage{appendix}
\usepackage[T1]{fontenc}
\usepackage{lmodern}
\usepackage[dvipsnames,x11names]{xcolor}
\usepackage[colorlinks=true,linkcolor=NavyBlue,citecolor=ForestGreen,urlcolor=NavyBlue]{hyperref}
\usepackage[sort&compress]{natbib}
\usepackage{lipsum}
\usepackage{morefloats}
\usepackage[pdf]{pstricks}
\usepackage{amsmath}
\usepackage{amssymb}
\usepackage{amsfonts}
\usepackage{rotating}
\usepackage{cancel}
\usepackage{mathtools}
\usepackage{bbm}
\usepackage{dsfont}
\usepackage{bbold}
\usepackage{multirow}
\usepackage{ulem}
\usepackage{physics}
\usepackage{orcidlink}
\usepackage{colortbl}

\def\xslash{x\!\!\!\slash }

\def\vel{\left|}
\def\ver{\right|}

\begin{document}

\title{Analysis of the $Z_b(10650)$ state based on electromagnetic properties}

\author{Ula\c{s}~\"{O}zdem\orcidlink{0000-0002-1907-2894}
}%
\email[]{ulasozdem@aydin.edu.tr }
\affiliation{Health Services Vocational School of Higher Education, Istanbul Aydin University, Sefakoy-Kucukcekmece, 34295 Istanbul, T\"{u}rkiye}

\date{\today}
 
\begin{abstract}
In this study, the magnetic and quadrupole moments of the $Z_b(10650)$ state are determined using the compact diquark-antidiquark interpolating current through the QCD light-cone sum rule. The values that are obtained as a result of the analysis are as follows: $\mu_{Z_b} =  2.35^{+0.34}_{-0.33}~\mu_N$ and $\mathcal{D}_{Z_b} =(1.82^{+0.35}_{-0.31})\times 10^{-2}~\mbox{fm}^2$. Examining the results obtained, it can be seen that the magnetic moments are large enough to be measured experimentally, while the quadrupole moment is obtained as a small but non-zero value, corresponding to a prolate charge distribution. The magnetic moment is the leading-order response of a bound system to a weak external magnetic field. It therefore provides an excellent platform to probe the internal structures of hadrons governed by the quark-gluon dynamics of QCD.
\end{abstract}
\keywords{Hidden-bottom tetraquarks, diquark-antidiquark picture, electromagnetic multipole moments, QCD light-cone sum rules}

\maketitle

\section{Motivation} \label{section1} 

Over the past decade, a series of heavy quarkonium-like states, known as the XYZ states, have been discovered through ongoing experimental efforts. Charged states such as $Z_c(3900)$, $Z_c(4430)$ and $Z_c(4020)$ provide strong evidence for the existence of exotic hadrons since the presence of light quarks explains their non-zero electric charge. These electrically charged particles have hidden-heavy flavor (hidden-charm or hidden-bottom), excluding the pure $Q\bar Q$, opening a new era in hadron physics. 
In 2011, two charged bottomonium-like states, 
$Z_b (10610)$ and $Z_b(10650)$, have been reported by the Belle Collaboration in the processes $\Upsilon(5S) \rightarrow \pi\pi \Upsilon(nS)$, 
and $\Upsilon(5S) \rightarrow \pi\pi h_b (kP )$~\cite{Belle:2011aa} with n = 1, 2, 3 and k = 1, 2. The masses and widths of these states were measured as
             \begin{eqnarray}
M_{Z_b (10610)}&=10607.2 \pm 2~\mbox{MeV}, ~~~~
              \Gamma_{Z_b (10610)}= 18.4 \pm 2.4~\mbox{MeV}, \\
M_{Z_b(10650)}&=10652.2 \pm 1.5~\mbox{MeV},~~
              \Gamma_{Z_b(10650)}= 11.5 \pm 2.2~\mbox{MeV}.
             \end{eqnarray}        
The neutral partner of the $Z_b^0(10610)$ was also observed in the $\Upsilon(5S) \to \Upsilon(nS) \pi \pi$ decay by the Belle Collaboration \cite{Belle:2013urd}.             
The quantum numbers $I^G(J^P) = 1^+ (1^+)$ are favored by the analysis of the angular distribution. Both $Z_b (10610)$ and $Z_b(10650)$ states are members of a family of charged hidden-bottom states. These states have attracted the attention of many theoretical groups since they were the first charged bottomonium-like states observed and because they are very close to the thresholds of $B \bar B^*$ and $B^* \bar B^*$. Various models and approaches have been used to study the spectroscopic parameters and decays of both states, including compact tetraquark states, molecular states, threshold cusps, re-scattering effects, etc. It is a challenge to understand these new charmonium/bottomonium-like states as exotic since it is relatively easy to reproduce the properties of these states using the models mentioned above.   Many comprehensive reviews of this topic can be found in the literature~\cite{Faccini:2012pj,Esposito:2014rxa,Chen:2016qju,Ali:2017jda,Esposito:2016noz,Olsen:2017bmm,Lebed:2016hpi,Guo:2017jvc,Brambilla:2019esw,Liu:2019zoy, Agaev:2020zad, Dong:2021juy,Chen:2015ata,Meng:2022ozq,Chen:2022asf}.  
All possible configurations considered in the various studies yielded mass and decay width determinations in agreement with experimental observations, which indicates that other properties of this state need to be further investigated to shed light on its underlying structure and reach a definitive conclusion. 
Therefore, to elucidate the internal organization of these states, it is also important to study the decay channels, such as weak, strong, and electromagnetic, along with the spectroscopic parameters of these states.

The magnetic and quadrupole moments are other intrinsic parameters of hadrons that may contain important information about their quark-gluon organization and underlying dynamics. Such a study would therefore deepen our knowledge of tetraquarks and help us understand the underlying dynamics that govern how they form, as well as the geometric shapes of these tetraquarks. The magnetic and quadrupole moments of the hidden-heavy tetraquarks were extracted in several studies~\cite{Ozdem:2017exj,Ozdem:2017jqh,Ozdem:2022kck,Ozdem:2021yvo,Ozdem:2021hka,Ozdem:2023frj,Xu:2020qtg,Wang:2017dce,Xu:2020evn,Wang:2023vtx}. 
In our study, the magnetic and quadrupole moments of the $Z_b(10650)$ state (for brevity, hereinafter
often referred to as $Z_b$)  are determined using the compact diquark-antidiquark interpolating current through the QCD light-cone sum rule. In the QCD light-cone sum rule method~\cite{Chernyak:1990ag, Braun:1988qv, Balitsky:1989ry}, we describe the correlation function in two different representations: One is based on the hadronic degrees of freedom and is called the hadronic representation, and the other is based on the quark-gluon degrees of freedom and is called the QCD representation.  Double Borel transformations on $p^2$ and $(p+q)^2$ are then applied to both representations to suppress the contributions of the higher states and the continuum. The quark-hadron duality ansatz is also carried out to further suppress the contributions of the higher states and the continuum, and to enhance the ground state contribution. The magnetic and quadrupole moments are obtained by matching the coefficients of the same Lorentz structures of both representations of the correlation function.

This article is organized as follows. After the introduction, we introduce our theoretical framework explicitly in section \ref{section2}. The numerical results and conclusions for the magnetic and quadrupole moments of the $Z_b$ state are presented in section \ref{section3}. Finally, the manuscript ends with a summary in section \ref{section4}.
Explicit expressions for the magnetic moment of the $Z_b$ state and the photon distribution amplitudes are listed in the appendices \ref{appenda} and \ref{appendb}, respectively.

\section{Electromagnetic multipole moments of the $Z_b(10650)$ state from QCD light-cone sum rules} \label{section2} 
In the QCD light-cone sum rule, one initiates the calculations for the magnetic and quadrupole moments using the following correlation function: 
\begin{equation}
 \label{edmn01}
\Pi _{\mu \nu }(p,q)=i\int d^{4}xe^{ip\cdot x}\langle 0|\mathcal{T}\{J_{\mu}(x)
J_{\nu }^{\dagger }(0)\}|0\rangle_{\gamma}, 
\end{equation}%
where  $q$ is the momentum of the photon, $\gamma$ denotes the external electromagnetic background field. 
The four-quark current operator $J_{\mu(\nu)}(x)$ with spin-parity quantum numbers $J^{P}=1^{+}$ is written as
\begin{align}
\label{curr}
    J_{\mu}(x)&=\frac{\epsilon \tilde{\epsilon}}{\sqrt{2}}\Big\{
    [ { u^{bT}}(x) C \sigma_{\alpha\mu} \gamma_5 b^c(x)]
    [ \bar d^d(x)  \gamma^{\alpha} C { \bar b^{eT}}(x)] 
    -[ { u^{bT}}(x) C \gamma^{\alpha}  b^c(x)][ \bar d^d(x) \gamma_5  \sigma_{\alpha\mu} C  {\bar b^{eT}} (x)]
\Big\},
  \end{align}
where  $\epsilon =\epsilon _{abc}$, $\tilde{\epsilon}=\epsilon _{ade}$ with color indices the $a$, $b$, $c$, $d$, and  $e$; $C$ being the charge conjugation operator, and $\sigma_{\mu\nu}=\frac{i}{2}[\gamma_{\mu},\gamma_{\nu}]$. 
It should be noted that magnetic and quadrupole moment calculations have been done under the assumption that the $Z_b$ state can also be in the $B^* \bar B^*$ molecular configuration, however, these results are not reported in the manuscript because a reliable sum rule could not be obtained.

In the hadronic representation, a complete set of hadronic states with the same quantum numbers as the state of interest and the corresponding interpolating current is plugged into the correlation function, which gives the hadronic representation in terms of the electromagnetic multipole moments as
 \begin{align}
\label{edmn04}
\Pi_{\mu\nu}^{Had} (p,q) &= {\frac{\langle 0 \mid J_\mu (x) \mid
Z_{b}(p, \varepsilon^i) \rangle}{p^2 - m_{Z_{b}}^2}} 
\langle Z_{b}(p, \varepsilon^i) \mid Z_{b}(p+q, \varepsilon^f) \rangle_\gamma 
 \frac{\langle Z_{b}(p+q,\varepsilon^f) \mid {J^\dagger}_\nu (0) \mid 0 \rangle}{(p+q)^2 - m_{Z_{b}}^2} 
+\mbox{higher states}.
\end{align}
 
The matrix elements $\langle 0 \mid J_\mu(x) \mid Z_{b}(p,\varepsilon^i) \rangle$, $\langle Z_{b}(p+q,\varepsilon^f) \mid {J^\dagger}_\nu (0) \mid 0 \rangle$ and $\langle Z_{b}(p,\varepsilon^i) \mid Z_{b} (p+q,\varepsilon^{f})\rangle_\gamma$ have been described regarding hadronic parameters such as residues, polarization vectors, and form factors, by the following expressions
\begin{align}
\langle 0 \mid J_\mu(x) \mid Z_{b}(p,\varepsilon^i) \rangle &= \lambda_{Z_{b}} \varepsilon_\mu^i\,,\\
\langle Z_{b}(p+q,\varepsilon^f) \mid {J^\dagger}_\nu (0) \mid 0 \rangle  &= \lambda_{Z_{b}} \varepsilon_\nu^{*f}\,,\\
\langle Z_{b}(p,\varepsilon^f) \mid  Z_{b} (p+q,\varepsilon^{i})\rangle_\gamma &= - \varepsilon^\gamma (\varepsilon^{i})^\alpha (\varepsilon^{f})^\beta 
 \bigg\{ G_1(Q^2) (2p+q)_\gamma ~g_{\alpha\beta} 
+ G_2(Q^2) ( g_{\gamma\beta}~ q_\alpha -  g_{\gamma\alpha}~ q_\beta) 
\nonumber\\ &- \frac{1}{2 m_{Z_{b}}^2} G_3(Q^2)~ (2p+q)_\gamma 
q_\alpha q_\beta  \bigg\},\label{edmn06}
\end{align}
where $\varepsilon^\gamma$ represents the photon's polarization vector, while $\varepsilon^{i}$ and $\varepsilon^{f}$ represent the polarization vectors of the initial and final $Z_b$ state, respectively.

Using the Eqs~(\ref{edmn04})-(\ref{edmn06}), the hadronic representation of the correlation function is,
\begin{align}
\label{edmn09}
 \Pi_{\mu\nu}^{Had}(p,q) &=  \frac{\varepsilon_\rho \, \lambda_{Z_b}^2}{ [m_{Z_b}^2 - (p+q)^2][m_{Z_b}^2 - p^2]}
 \bigg\{G_1(Q^2)  (2p+q)_\rho\bigg[g_{\mu\nu}-\frac{p_\mu p_\nu}{m_{Z_b}^2}
 -\frac{(p+q)_\mu (p+q)_\nu}{m_{Z_b}^2} \nonumber\\
 &+\frac{(p+q)_\mu p_\nu}{2m_{Z_b}^4} (Q^2+2m_{Z_b}^2)
 \bigg]
 + G_2 (Q^2) \bigg[q_\mu g_{\rho\nu} - q_\nu g_{\rho\mu} -
\frac{p_\nu}{m_{Z_b}^2}  \big(q_\mu p_\rho - \frac{1}{2}
Q^2 g_{\mu\rho}\big) \nonumber\\
&  
+
\frac{(p+q)_\mu}{m_{Z_b}^2}  \big(q_\nu (p+q)_\rho+ \frac{1}{2}
Q^2 g_{\nu\rho}\big)
-  
\frac{(p+q)_\mu p_\nu p_\rho}{m_{Z_b}^4} \, Q^2
\bigg]\nonumber\\
&
-\frac{G_3(Q^2)}{m_{Z_b}^2}(2p+q)_\rho \bigg[
q_\mu q_\nu -\frac{p_\mu q_\nu}{2 m_{Z_b}^2} Q^2 +\frac{(p+q)_\mu q_\nu}{2 m_{Z_b}^2} Q^2
-\frac{(p+q)_\mu q_\nu}{4 m_{Z_b}^4} Q^4\bigg]
\bigg\}\,.
\end{align}

 The magnetic and quadrupole moments of hadrons are related to their magnetic ($F_M(Q^2)$) and quadrupole ($F_{\mathcal D} (Q^2)$)  form factors. 
The form factors $F_M(Q^2)$ and $F_{\mathcal D} (Q^2)$, which are more directly accessible experimentally, are given by the form factors $G_i(Q^2)$
\begin{align}
\label{edmn07}
&F_M(Q^2) = G_2(Q^2)\,,\nonumber \\
&F_{\mathcal D}(Q^2) = G_1(Q^2)-G_2(Q^2)+(1+\beta) G_3(Q^2)\,,
\end{align}
where  $\beta=Q^2/4 m_{Z_b}^2$ with $Q^2=-q^2$.  At zero momentum transfer, the magnetic ($\mu_{Z_b}$) and quadrupole ($\mathcal {D}_{Z_b}$) moments are described by the form factors $F_M(Q^2=0)$ and $F_{\mathcal D}(Q^2=0)$ as follows
\begin{align}
\label{edmn08}
& \mu_{Z_b}= \frac{e}{2 m_{Z_b}}  F_M(Q^2=0) \,, \nonumber\\
& \mathcal {D}_{Z_b}  = \frac{e}{m_{Z_b}^2} F_{\cal D}(Q^2=0)\,.
\end{align}

The QCD representation of the evaluations requires the use of the interpolating field explicitly in the correlation function, Eq. (\ref{edmn01}). This is followed by all the possible contractions of the quark operators according to Wick's theorem, which turns the result into the one that contains the quark propagators in the form of
\begin{align}
\label{neweq}
\Pi _{\mu \nu }^{\mathrm{QCD}}(p,q)&=i\,\frac{ \epsilon \tilde{\epsilon} \epsilon^{\prime} \tilde{\epsilon}^{\prime}}{2}
\int d^{4}xe^{ipx} \langle 0 | \Big\{ 
\mathrm{Tr}\Big[\gamma^{\alpha}{\tilde S}_{b}^{e^{\prime }e}(-x)\gamma ^{\beta}  S_{d}^{d^{\prime }d}(-x)\Big] 
\mathrm{Tr}\Big[\sigma_{\mu\alpha}\gamma _{5 }{S}_{b}^{cc^{\prime }}(x)\gamma _{5}\sigma_{\nu\beta}\tilde S_{u}^{bb^{\prime }}(x)\Big] \notag \\
&-\mathrm{Tr}\Big[ \gamma^{\alpha}{\tilde S}_{b}^{e^{\prime }e}(-x)\sigma_{\nu\beta}\gamma _{5}S_{d}^{d^{\prime }d}(-x)\Big]   
\mathrm{Tr}\Big[ \sigma_{\mu\alpha}\gamma_{5 }
  {S}_{b}^{cc^{\prime }}(x)\gamma^{\beta}\tilde S_{u}^{bb^{\prime }}(x)\Big] 
  \nonumber\\
&
-\mathrm{Tr}\Big[ \gamma _{5}\sigma_{\mu\alpha}{\tilde S}_{b}^{e^{\prime }e}(-x)
 \gamma^{\beta }S_{d}^{d^{\prime }d}(-x)\Big]    
\mathrm{Tr}\Big[ \gamma^{\alpha}{S}_{b}^{cc^{\prime }}(x)\gamma_{5}\sigma_{\nu\beta}\tilde S_{u}^{bb^{\prime }}(x)\Big] \nonumber \\
&+\mathrm{Tr}\Big[\gamma_{5 }\sigma_{\mu\alpha}{\tilde S}_{b}^{e^{\prime }e}(-x)\sigma_{\nu\beta}\gamma _{5}S_{d}^{d^{\prime }d}(-x)\Big]  
\mathrm{Tr}\Big[\gamma^{\alpha} {S}_{b}^{cc^{\prime }}(x) \gamma^{\beta}\tilde S_{u}^{bb^{\prime }}(x)\Big]
 \Big\}| 0 \rangle_\gamma,
\end{align} 
where $S_{b}(x)$ and $S_{q}(x)$ are the propagators for heavy and light quarks, respectively. The explicit formulas of these propagators are given in the following form~\cite{Yang:1993bp, Belyaev:1985wza}

\begin{align}
\label{edmn12}
S_{q}(x) &= S_q^{free}(x)
- \frac{\langle \bar qq \rangle }{12} \Big(1-i\frac{m_{q} \xslash}{4}   \Big)
- \frac{\langle \bar q \sigma.G q \rangle }{192}x^2 
 \Big(1-i\frac{m_{q} \xslash}{6}   \Big)
-\frac {i g_s }{32 \pi^2 x^2} ~G^{\mu \nu} (x) \bigg[\rlap/{x}
\sigma_{\mu \nu} +  \sigma_{\mu \nu} \rlap/{x}
 \bigg],
 \\
S_{b}(x)&=S_b^{free}(x)
-\frac{g_{s}m_{b}}{16\pi ^{2}} \int_0^1 dv\, G^{\mu \nu }(vx)\Bigg[ (\sigma _{\mu \nu }{\xslash}
  +{\xslash}\sigma _{\mu \nu }) 
  \frac{K_{1}\Big( m_{b}\sqrt{-x^{2}}\Big) }{\sqrt{-x^{2}}}
+2\sigma_{\mu \nu }K_{0}\Big( m_{b}\sqrt{-x^{2}}\Big)\Bigg].
\label{edmn13}
\end{align}%
where 
\begin{align}
S_q^{free}(x) &=\frac{1}{2 \pi^2 x^2}\Big( i \frac{{\xslash}}{x^{2}}-\frac{m_{q}}{2 } \Big),\\
\nonumber\\
S_b^{free}(x) &= \frac{m_{b}^{2}}{4 \pi^{2}} \Bigg[ \frac{K_{1}\Big(m_{b}\sqrt{-x^{2}}\Big) }{\sqrt{-x^{2}}}
+i\frac{{\xslash}~K_{2}\Big( m_{b}\sqrt{-x^{2}}\Big)}
{(\sqrt{-x^{2}})^{2}}\Bigg].
\end{align}


The correlation functions specified in Eq.~(\ref{neweq}) contain both perturbative and non-perturbative contributions from short and long distances, respectively. To derive formulas for the perturbative contributions, i.e. when the photon is radiated at a short-distance, it is sufficient to replace one of the light or heavy propagators in Eq.~(\ref{neweq}) by the following
\begin{align}
\label{free}
S^{free}(x) \rightarrow \int d^4z\, S^{free} (x-z)\,\rlap/{\!A}(z)\, S^{free} (z)\,,
\end{align}
where the three surviving propagators in eq. (\ref{neweq}) are considered to be the free ones.
To derive the formulas for the nonperturbative contributions, i.e.  where the photon is radiated at a long distance, replace one of the light quark propagators in the correlation function given in Eq. (\ref{neweq}) as follows
\begin{align}
\label{edmn14}
S_{\mu\nu}^{ab}(x) \rightarrow -\frac{1}{4} \big[\bar{q}^a(x) \Gamma_i q^b(0)\big]\big(\Gamma_i\big)_{\mu\nu},
\end{align}
 where  the remaining propagators in Eq. (\ref{neweq}) are taken into account as full propagators, and  
 $\Gamma_i = \textbf{1}, \gamma_5, \gamma_\mu, i\gamma_5 \gamma_\mu, \sigma_{\mu\nu}/2$. 
 When a photon interacts nonperturbatively with light-quark fields,   the matrix elements of the nonlocal operators $\langle \gamma(q)\vel \bar{q}(x) \Gamma_i q(0) \ver 0\rangle$ and $\langle \gamma(q)\vel \bar{q}(x) \Gamma_i G_{\mu\nu}q(0) \ver 0\rangle$  appear between the vacuum and the photon state,  expressed in terms of photon distribution amplitudes (DAs) (for details see Ref. \cite{Ball:2002ps}).  Eqs.~(\ref {neweq})-(\ref {edmn14}) are used for evaluation the QCD representation of the correlation function.  The expression of the correlation function in the x-space is then transferred to the momentum space by means of the Fourier transform.

The physical quantities that we search for in this section, namely the magnetic and quadrupole moments of the $Z_b$ state, are obtained by matching the coefficients of the same Lorentz structures, $(\varepsilon.p) (p_\mu q_\nu- p_\nu q_\mu)$ and $(\varepsilon.p) q_\mu q_\nu$, acquired in both the hadronic and QCD representations. These matches are written as
\begin{align}
 \mu_{Z_{b}}  &= \frac{e^{\frac{m_{Z_{b}}^2}{M^2}}}{ \lambda_{Z_{b}}^2}  \,\, \Delta_1^{QCD}(M^2,s_0),\label{sonj1} \\
 \mathcal{D}_{Z_b}  &= \frac{ m_{Z_{b}}^2 e^{\frac{m_{Z_{b}}^2}{M^2}}}{\lambda_{Z_{b}}^2} \,\, \Delta_2^{QCD}(M^2,s_0), \label{sonj2} 
 \end{align}
where $M^2$ and $s_0$ represent the Borel mass the continuum threshold parameter, respectively. For brevity, since the $\Delta_1^{QCD}(M^2,s_0)$ and $\Delta_2^{QCD}(M^2,s_0)$ functions have a similar form, only the explicit expressions of the $\Delta_1^{QCD}(M^2,s_0)$ function are given in the Appendix \ref{appenda}.

\section{Numerical analysis and conclusion} \label{section3}

In this section, we conduct a numerical analysis of the QCD light-cone sum rules for the magnetic and quadrupole moments of the $Z_b$ tetraquark state, which were derived in the previous section. In the numerical analysis following input parameters are used:  $m_u=m_d=0$, 
$m_b = 4.18^{+0.03}_{-0.02}\,$GeV,  $m_{Z_b} = 10652.2\pm 1.5\,$MeV~\cite{ParticleDataGroup:2022pth}, $\lambda_{Z_b}=(2.12 \pm 0.31) \times 10^{-1}$~GeV$^5$ \cite{Wang:2019mxn}, 
$\langle \bar uu\rangle $ =  
$\langle \bar dd\rangle$=$(-0.24\pm0.01)^3\,$GeV$^3$  \cite{Ioffe:2005ym},   
and $\langle g_s^2G^2\rangle = 0.88~ $GeV$^4$~\cite{Matheus:2006xi}. 
The photon DAs and corresponding parameters can be found in Appendix \ref{appendb}.

The sum rules are obtained for the magnetic and quadrupole moments given in Eqs. (\ref{sonj1}) and (\ref{sonj2}), which also depend on the Borel and continuum threshold parameters $M^2$ and $s_0$. The choice of working regions for $M^2$ and $s_0$ has to fulfill standard restrictions imposed on the pole contribution (PC) and the convergence of the operator product expansion (OPE).  In order to quantify these constraints, it is appropriate to use the expressions
\begin{align}
 \mbox{PC} &=\frac{\Delta (M^2,s_0)}{\Delta (M^2,\infty)} \geq  30\%,
\\
\nonumber\\
 \mbox{OPE Convergence} &=\frac{\Delta^{\scriptsize{\mbox{Dim7}}} (M^2,s_0)}{\Delta (M^2,s_0)}\leq  5\%,
 \end{align}
 where $\Delta^{\mathrm{Dim7}} (M^2,s_0)$ is the contribution of the highest dimensional term in the OPE. 
Because of the aforementioned constraints, the obtained working regions for $M^2$ and $s_0$ are provided as follows: $11~\mbox{GeV}^2 \leq M^2 \leq 15~\mbox{GeV}^2$ and $121~\mbox{GeV}^2 \leq s_0 \leq 125~\mbox{GeV}^2$. 
 Results obtained for the PC values and the OPE convergence in the working intervals obtained for auxiliary parameters: $31.2\% \leq  \mbox{PC}  \leq 53.4\%$ and $\mbox{OPE} \leq 3.69 \%$.  From these values, we can see that the working regions determined for $M^2$ and $s_0$ meet the above requirements.
For the sake of completeness, it is worth examining how the magnetic and quadrupole moments depend on $M^2$ for different values of $s_0$.  From fig. 1 it is seen that the magnetic and quadrupole moments exhibit relatively mild dependence on the variation of $M^2$ within its working region.

\begin{widetext}

\begin{figure}[htb]
\label{Msqfig}
\centering
 \includegraphics[width=0.47\textwidth]{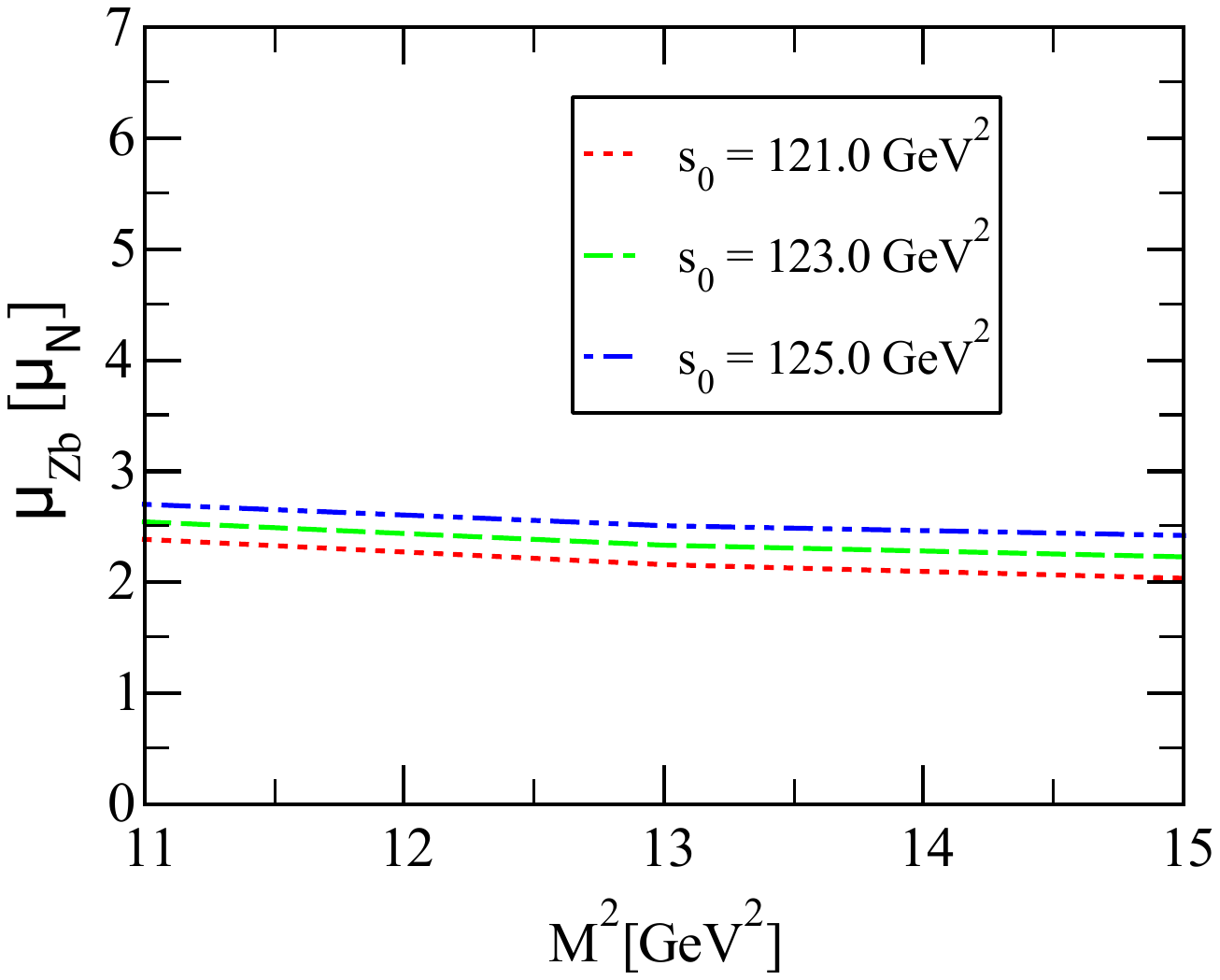}
 \includegraphics[width=0.47\textwidth]{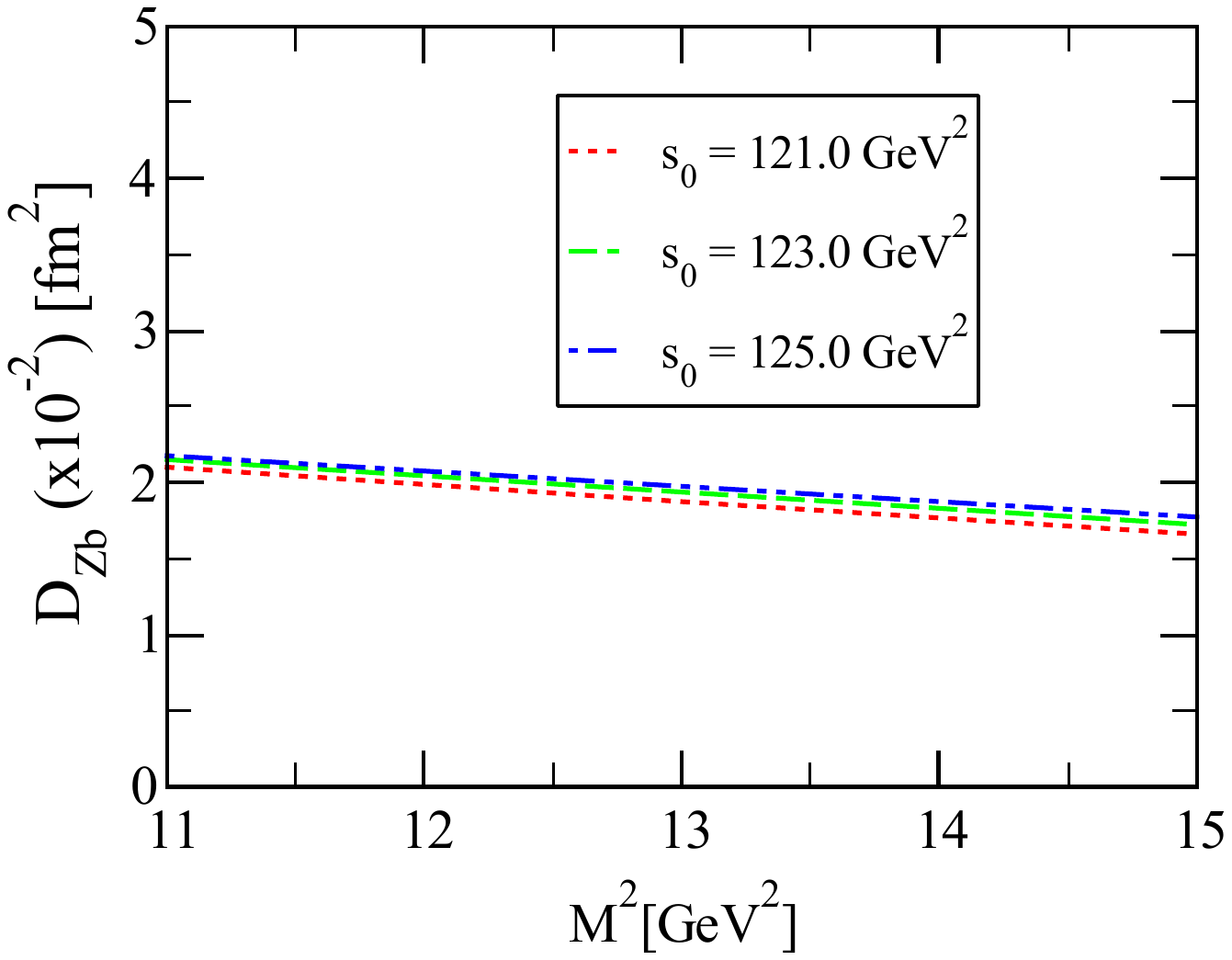}
 \caption{ Magnetic and quadrupole moments of the $Z_b$ state versus $M^2$ for various $s_0$ values.}
  \end{figure}

  \end{widetext}

The magnetic and quadrupole moment results, taking into account the uncertainties in the input parameters and the variation in the $M^{2}$ and $s_0$ working regions, are given below:
\begin{align}
 &\mu_{Z_b} = 2.35^{+0.34}_{-0.33}~\mu_N\,,\\
 \nonumber\\
 &\mathcal{D}_{Z_b} =(1.82^{+0.35}_{-0.31})\times 10^{-2}~\mbox{fm}^2.
\end{align}

The order of the numerical results of the magnetic moments can also give an insight into the experimental measurement of them. By examining the magnetic moment result, one can assume that the magnetic moment of $Z_b$ is large enough to be measured in future experiments with the increased luminosity. In the case of the quadrupole moment, we get a non-zero but small value for the $Z_b$ state, indicating a non-spherical charge distribution. It is well known that the sign of the quadrupole moment contains information about the geometric shape of the hadron under study. For the $Z_b$ state, the quadrupole moment has a positive sign, corresponding to a prolate charge distribution. 
To the best of our knowledge, this is the first study of the magnetic and quadrupole moments of the $Z_b(10650)$ state, so there are no theoretical or experimental results to compare.  
However, to give an idea of the results obtained, a comparison can be made with the magnetic moment results obtained by the QCD light-cone sum rules method for the hidden-charm and hidden-bottom tetraquark states in the molecular and compact diquark-antidiquark pictures.  In Ref.~\cite{Ozdem:2017exj}, the magnetic moment of the $Z_b(10610)$ state in the molecular and compact diquark-antidiquark  pictures was extracted by means of the QCD light-cone sum rules.
The result obtained are $\mu_{Z_b(10610)}=1.73 \pm 0.63~\mu_N$ and $\mu_{Z_b(10610)}=1.59 \pm 0.58~\mu_N$ for the compact diquark-antidiquark and molecular configuration, respectively. Based on the presented results, it can be seen that the magnetic moment of the $Z_b(10650)$ state is of the same order as that of the $Z_b(10610)$ state. 
%
%
In Refs. \cite{Ozdem:2017jqh,Ozdem:2022kck,Ozdem:2021yvo,Ozdem:2021hka,Ozdem:2023frj}, the authors systematically investigated the magnetic moments of various hidden-charm tetraquark states using the molecular and compact diquark-antidiquark configurations within the framework of the QCD light-cone sum rules method. The magnetic  moment values obtained for these states are roughly of the order of $(0.5-1.0)~\mu_N$.
In Refs. \cite{Xu:2020qtg,Wang:2017dce}, the magnetic moment of the $Z_c(3900)$ state in the molecular and compact diquark-antidiquark pictures was obtained by the QCD sum rule method in an external weak electromagnetic field, and they obtained $\mu_{Z_c(3900)}=0.19^{+0.04}_{-0.01}~\mu_N$ and $\mu_{Z_c(3900)}=0.47^{+0.27}_{-0.22}~\mu_N$, for the molecular and compact diquark-antidiquark pictures, respectively. 
In Ref. \cite{Xu:2020evn}, the authors have employed the QCD sum rule method in an external weak electromagnetic field to calculate the magnetic moment of the $Z_{cs}(3985)$ state in the molecular picture, and they obtained $\mu_{Z_{cs}(3985)}=0.18^{+0.16}_{-0.09}~\mu_N$. In Ref. \cite{Wang:2023vtx}, the authors have applied the multiquark color flux-tube model to extract the magnetic moments of the $Z_{cs}(4000)$ and $Z_{cs}(4220)$  states in the compact diquark-antidiquark picture, and they obtained $\mu_{Z_{cs}(4000)}=0.73~\mu_N$ and $\mu_{Z_{cs}(4220)}=0.64~\mu_N$.
As can be seen from these values, the results obtained for the hidden-charm and hidden-bottom states are quite different from each other.

It is useful to consider the individual quark sector contributions to the magnetic and quadrupole moments to gain a deeper understanding of the underlying quark-gluon dynamics. This can be achieved by selecting the appropriate charge factors $e_u$, $e_d$, and $e_b$. When it has been done magnetic moment we see that the terms proportional to the $e_u$ contribute about $\%67$ to the total results, $e_d$ about $\%33$, and $e_b$ is zero.  In the case of the quadrupole moment, we observe that the proportional terms of $e_u$ contribute about $\%66$ to the total results, $e_d$ contributes about $\%34$, and $e_b$ is zero. A closer look reveals that the missing $e_b$ contribution is because the terms containing this term exactly cancel each other out.

\section{Summary and outlook
} \label{section4}

In the present article, we have explored the magnetic and quadrupole moments of the $Z_b(10650)$ state with the spin-parity $J^P=1^+$  in the framework of the QCD light-cone sum rule by modeling this state as a  compact diquark-antidiquark configuration. 
Examining the results obtained, it can be seen that the magnetic moments are large enough to be measured experimentally, while the quadrupole moment is obtained as a small but non-zero value, corresponding to a prolate charge distribution.  Individual quark sector contributions to the magnetic and quadrupole moments have also been analyzed, and it has been found that only light quarks contribute to these quantities. Our predictions should also be checked through other phenomenological methods as well. The results of this study on the magnetic and quadrupole moments of the $Z_b(10650)$ state can be used in future experimental studies of the exotic states. 

\section{Acknowledgements}
The author thanks A. \"{O}zpineci for useful discussions, comments, and suggestions.

 \appendix
\begin{widetext}

\section{ The explicit expressions of the $\Delta_1^{QCD}(M^2, s_0)$} \label{appenda}
The explicit forms of the $\Delta_1^{QCD}(M^2, s_0)$ functions appearing in the above sum rules are

\begin{align}
 \Delta_1^{QCD}(M^2,s_0)&=\frac {27 (e_d - e_u)} {1310720 \pi^5}\Bigg[
   I[0, 5, 3, 1] - 3 I[0, 5, 3, 2] + 3 I[0, 5, 3, 3] - 
    I[0, 5, 3, 4] - 3 I[0, 5, 4, 1] \nonumber\\
    &+ 6 I[0, 5, 4, 2] - 3 I[0, 5, 4, 3] + 3 I[0, 5, 5, 1] - 3 I[0, 5, 5, 2] - 
    I[0, 5, 6, 1]\Bigg]\nonumber\\
        & + \frac {m_b^2 (e_d - e_u) } {65536 \pi^5}\Bigg[
    I[0, 4, 2, 2] - 2 I[0, 4, 2, 3] + I[0, 4, 2, 4] - 
     2 I[0, 4, 3, 2] + 2 I[0, 4, 3, 3] + 
     I[0, 4, 4, 2] \Bigg] \nonumber \\
  &  -\frac {m_b \langle g_s^2 G^2\rangle \langle \bar q q \rangle } {7077888 \pi^3} \Bigg[e_d \Big (-92 I_ 4[\mathcal S] + 60 I_ 4[\mathcal T_1] + 
       159 I_ 4[\mathcal T_ 2] - 21 I_ 4[\mathcal T_ 3] + 
       78 I_ 4[\mathcal T_ 4] + 11 I_ 4[\mathcal {\tilde S}]\Big)
       - 
    e_u \Big (117 I_ 3[\mathcal S] \nonumber\\
       & + 60 I_ 3[\mathcal T_ 1] + 
       159 I_ 3[\mathcal T_ 2] - 21 I_ 3[\mathcal T_ 3] + 
       78 I_ 3[\mathcal T_ 4] + 160 I_ 3[\mathcal {\tilde S}]\Big)- 
    192 (e_d - e_u) I_6[h_ {\gamma} ]\Bigg] I[0, 1, 3, 0]\nonumber\\
       & +\frac {f_ {3\gamma} \langle g_s^2 G^2\rangle } {28311552 \pi^3} \Bigg[-128  m_b^2 (e_d - e_u)  (52 I[0, 1, 2, 0] - 
     3 I[0, 1, 3, 
       0]) I_ 6[\psi_ {\gamma}^{\nu}] - \Big (27 e_u I_ 1[\mathcal A] - 27 e_d I_ 2[\mathcal A] \nonumber\\
&+ 
     128 (3 e_d - e_u) I_ 6[\psi_ {\gamma}^{\nu}]\Big)\Bigg] \nonumber
 \end{align}
\begin{align}
       &
       +\frac {m_b \langle \bar q q \rangle} {393216 \pi^3} \Bigg[ 4 \Bigg (-e_u \Big (5 I_ 3[\mathcal S] - 23 I_ 3[\mathcal T_ 1] - 
      23 I_ 3[\mathcal T_ 2] + I_ 3[\mathcal {\tilde S}]\Big) + 
   e_d  \Big (23 I_ 4[\mathcal T_ 1] + 23 I_ 4[\mathcal T_ 2] + 
       22  I_ 4[\mathcal {\tilde S}]\Big)\Bigg)\nonumber\\
&\times  I[0, 3, 4, 0]
       -3 \Bigg (3 e_u \Big (I_ 3[\mathcal S] + I_ 3[\mathcal T_ 1] + 
     I_ 3[\mathcal T_ 2] + I_ 3[\mathcal {\tilde S}]\Big)
     -e_d \Big (2  I_ 4[\mathcal S] - 3 I_ 4[\mathcal T_ 1] - 
   3 I_ 4[\mathcal T_ 2] + I_ 4[\mathcal {\tilde S}]\Big)\nonumber\\
   &+96 (e_d - e_u) I_ 6[h_ {\gamma}]\Bigg) I[0, 3, 5, 0]
 \Bigg] \nonumber\\
& +\frac { f_{3\gamma}} {6291456\pi^3} \Bigg[4 \Big (512 m_b^2 (e_d - e_u)   I_ 6[\psi_ {\gamma}^{\nu}] I[0, 3, 4, 
      0] - 3  \big (-e_u I_ 1[\mathcal V] + 
       9 e_d I_ 2[\mathcal V]\big) I[0, 4, 5, 0]\Big)  \nonumber\\
 &+ 
 9  \Big (e_u I_ 1[\mathcal V] - e_d I_ 2[\mathcal V]- 
    64 (3 e_d - e_u) I_ 6[\psi_ {\gamma}^{\nu}]\Big) I[0, 4, 6, 0]
      \Bigg], 
\end{align}
where the functions~$I[n,m,l,k]$, $I_1[\mathcal{F}]$,~$I_2[\mathcal{F}]$,~$I_3[\mathcal{F}]$,~$I_4[\mathcal{F}]$,
~$I_5[\mathcal{F}]$, and ~$I_6[\mathcal{F}]$ are
defined as:
\begin{align}
 I[n,m,l,k]&= \int_{4 m_c^2}^{s_0} ds \int_{0}^1 dt \int_{0}^1 dw~ e^{-s/M^2}~
 s^n\,(s-4\,m_c^2)^m\,t^l\,w^k,\nonumber\\
  I_1[\mathcal{F}]&=\int D_{\alpha_i} \int_0^1 dv~ \mathcal{F}(\alpha_{\bar q},\alpha_q,\alpha_g)
 \delta'(\alpha_ q +\bar v \alpha_g-u_0),\nonumber\\
  I_2[\mathcal{F}]&=\int D_{\alpha_i} \int_0^1 dv~ \mathcal{F}(\alpha_{\bar q},\alpha_q,\alpha_g)
 \delta'(\alpha_{\bar q}+ v \alpha_g-u_0),\nonumber\\
   I_3[\mathcal{F}]&=\int D_{\alpha_i} \int_0^1 dv~ \mathcal{F}(\alpha_{\bar q},\alpha_q,\alpha_g)
 \delta(\alpha_ q +\bar v \alpha_g-u_0),\nonumber\\
   I_4[\mathcal{F}]&=\int D_{\alpha_i} \int_0^1 dv~ \mathcal{F}(\alpha_{\bar q},\alpha_q,\alpha_g)
 \delta(\alpha_{\bar q}+ v \alpha_g-u_0),\nonumber\\
   I_5[\mathcal{F}]&=\int_0^1 du~ F(u)\delta'(u-u_0),\nonumber\\
 I_6[\mathcal{F}]&=\int_0^1 du~ F(u),
 \end{align}
 where $\mathcal{F}$ represents the corresponding photon DAs.

 \section{Photon Distribution Amplitudes and Wave Functions}\label{appendb}
In this appendix, we present the matrix elements $\langle \gamma(q)\vel \bar{q}(x) \Gamma_i q(0) \ver 0\rangle$  and $\langle \gamma(q)\vel \bar{q}(x) \Gamma_i G_{\mu\nu}q(0) \ver 0\rangle$ in terms of the photon DAs and wave functions of different twists. 
The expansion of the matrix element is an expansion in increasing twists of the DAs.
The twist of a DA is defined as the
dimension minus the spin of the operators contributing to a given DA. The DAs $\varphi_\gamma(u)$ have twist two, 
$\psi^v(u)$, $\psi^a(u)$, ${\cal A}(\alpha_i)$ and ${\cal V}(\alpha_i)$ have twist 3,  and $h_\gamma(u)$, $\mathbb{A}(u)$, ${\cal S}(\alpha_i)$, ${\cal{\tilde S}}(\alpha_i)$, ${\cal T}_1(\alpha_i)$, ${\cal T}_2(\alpha_i)$, ${\cal T}_3(\alpha_i)$  and ${\cal T}_4(\alpha_i)$ have twist 4.   The matrix elements $\langle \gamma(q)\vel \bar{q}(x) \Gamma_i q(0) \ver 0\rangle$  and $\langle \gamma(q)\vel \bar{q}(x) \Gamma_i G_{\mu\nu}q(0) \ver 0\rangle$ are parameterized in terms of the photon DAs as follows~\cite{Ball:2002ps}
\begin{eqnarray}
\label{esbs14}
&&\langle \gamma(q) \vert  \bar q(x) \gamma_\mu q(0) \vert 0 \rangle
= e_q f_{3 \gamma} \left(\varepsilon_\mu - q_\mu \frac{\varepsilon
x}{q x} \right) \int_0^1 du e^{i \bar u q x} \psi^v(u),
\nonumber \\
&&\langle \gamma(q) \vert \bar q(x) \gamma_\mu \gamma_5 q(0) \vert 0
\rangle  = - \frac{1}{4} e_q f_{3 \gamma} \epsilon_{\mu \nu \alpha
\beta } \varepsilon^\nu q^\alpha x^\beta \int_0^1 du e^{i \bar u q
x} \psi^a(u),
\nonumber \\
&&\langle \gamma(q) \vert  \bar q(x) \sigma_{\mu \nu} q(0) \vert  0
\rangle  = -i e_q \langle \bar q q \rangle (\varepsilon_\mu q_\nu - \varepsilon_\nu
q_\mu) \int_0^1 du e^{i \bar u qx} \left(\chi \varphi_\gamma(u) +
\frac{x^2}{16} \mathbb{A}  (u) \right) \nonumber \\ 
&&-\frac{i}{2(qx)}  e_q \langle \bar q q \rangle \left[x_\nu \left(\varepsilon_\mu - q_\mu
\frac{\varepsilon x}{qx}\right) - x_\mu \left(\varepsilon_\nu -
q_\nu \frac{\varepsilon x}{q x}\right) \right] \int_0^1 du e^{i \bar
u q x} h_\gamma(u),
\nonumber 
\end{eqnarray}
\begin{eqnarray}
&&\langle \gamma(q) | \bar q(x) g_s G_{\mu \nu} (v x) q(0) \vert 0
\rangle = -i e_q \langle \bar q q \rangle \left(\varepsilon_\mu q_\nu - \varepsilon_\nu
q_\mu \right) \int {\cal D}\alpha_i e^{i (\alpha_{\bar q} + v
\alpha_g) q x} {\cal S}(\alpha_i),
\nonumber \\
&&\langle \gamma(q) | \bar q(x) g_s \tilde G_{\mu \nu}(v
x) i \gamma_5  q(0) \vert 0 \rangle = -i e_q \langle \bar q q \rangle \left(\varepsilon_\mu q_\nu -
\varepsilon_\nu q_\mu \right) \int {\cal D}\alpha_i e^{i
(\alpha_{\bar q} + v \alpha_g) q x} \tilde {\cal S}(\alpha_i),
\nonumber \\
&&\langle \gamma(q) \vert \bar q(x) g_s \tilde G_{\mu \nu}(v x)
\gamma_\alpha \gamma_5 q(0) \vert 0 \rangle = e_q f_{3 \gamma}
q_\alpha (\varepsilon_\mu q_\nu - \varepsilon_\nu q_\mu) \int {\cal
D}\alpha_i e^{i (\alpha_{\bar q} + v \alpha_g) q x} {\cal
A}(\alpha_i),
\nonumber \\
&&\langle \gamma(q) \vert \bar q(x) g_s G_{\mu \nu}(v x) i
\gamma_\alpha q(0) \vert 0 \rangle = e_q f_{3 \gamma} q_\alpha
(\varepsilon_\mu q_\nu - \varepsilon_\nu q_\mu) \int {\cal
D}\alpha_i e^{i (\alpha_{\bar q} + v \alpha_g) q x} {\cal
V}(\alpha_i), \nonumber\\
&& \langle \gamma(q) \vert \bar q(x)
\sigma_{\alpha \beta} g_s G_{\mu \nu}(v x) q(0) \vert 0 \rangle  =
e_q \langle \bar q q \rangle \left\{
        \left[\left(\varepsilon_\mu - q_\mu \frac{\varepsilon x}{q x}\right)\left(g_{\alpha \nu} -
        \frac{1}{qx} (q_\alpha x_\nu + q_\nu x_\alpha)\right) \right. \right. q_\beta
\nonumber \\
 && -
         \left(\varepsilon_\mu - q_\mu \frac{\varepsilon x}{q x}\right)\left(g_{\beta \nu} -
        \frac{1}{qx} (q_\beta x_\nu + q_\nu x_\beta)\right) q_\alpha
-
         \left(\varepsilon_\nu - q_\nu \frac{\varepsilon x}{q x}\right)\left(g_{\alpha \mu} -
        \frac{1}{qx} (q_\alpha x_\mu + q_\mu x_\alpha)\right) q_\beta
\nonumber\\
 &&+
         \left. \left(\varepsilon_\nu - q_\nu \frac{\varepsilon x}{q.x}\right)\left( g_{\beta \mu} -
        \frac{1}{qx} (q_\beta x_\mu + q_\mu x_\beta)\right) q_\alpha \right]
   \int {\cal D}\alpha_i e^{i (\alpha_{\bar q} + v \alpha_g) qx} {\cal T}_1(\alpha_i)
\nonumber \\
 &&+
        \left[\left(\varepsilon_\alpha - q_\alpha \frac{\varepsilon x}{qx}\right)
        \left(g_{\mu \beta} - \frac{1}{qx}(q_\mu x_\beta + q_\beta x_\mu)\right) \right. q_\nu
\nonumber \\ &&-
         \left(\varepsilon_\alpha - q_\alpha \frac{\varepsilon x}{qx}\right)
        \left(g_{\nu \beta} - \frac{1}{qx}(q_\nu x_\beta + q_\beta x_\nu)\right)  q_\mu
\nonumber \\ && -
         \left(\varepsilon_\beta - q_\beta \frac{\varepsilon x}{qx}\right)
        \left(g_{\mu \alpha} - \frac{1}{qx}(q_\mu x_\alpha + q_\alpha x_\mu)\right) q_\nu
\nonumber \\ &&+
         \left. \left(\varepsilon_\beta - q_\beta \frac{\varepsilon x}{qx}\right)
        \left(g_{\nu \alpha} - \frac{1}{qx}(q_\nu x_\alpha + q_\alpha x_\nu) \right) q_\mu
        \right]      
    \int {\cal D} \alpha_i e^{i (\alpha_{\bar q} + v \alpha_g) qx} {\cal T}_2(\alpha_i)
\nonumber \\
&&+\frac{1}{qx} (q_\mu x_\nu - q_\nu x_\mu)
        (\varepsilon_\alpha q_\beta - \varepsilon_\beta q_\alpha)
    \int {\cal D} \alpha_i e^{i (\alpha_{\bar q} + v \alpha_g) qx} {\cal T}_3(\alpha_i)
\nonumber \\ &&+
        \left. \frac{1}{qx} (q_\alpha x_\beta - q_\beta x_\alpha)
        (\varepsilon_\mu q_\nu - \varepsilon_\nu q_\mu)
    \int {\cal D} \alpha_i e^{i (\alpha_{\bar q} + v \alpha_g) qx} {\cal T}_4(\alpha_i)
                        \right\},
\end{eqnarray}
where the integral measure ${\cal D} \alpha_i$ is written as

\begin{eqnarray}
\label{nolabel05}
\int {\cal D} \alpha_i = \int_0^1 d \alpha_{\bar q} \int_0^1 d
\alpha_q \int_0^1 d \alpha_g \delta(1-\alpha_{\bar
q}-\alpha_q-\alpha_g)~.
\end{eqnarray}

The explicit equations for the photon DAs with different twists are provided as

\begin{eqnarray}
\varphi_\gamma(u) &=& 6 u \bar u \left( 1 + \varphi_2(\mu)
C_2^{\frac{3}{2}}(u - \bar u) \right),
\nonumber \\
\psi^v(u) &=& 3 \left(3 (2 u - 1)^2 -1 \right)+\frac{3}{64} \left(15
w^V_\gamma - 5 w^A_\gamma\right)
                        \left(3 - 30 (2 u - 1)^2 + 35 (2 u -1)^4
                        \right),
\nonumber \\
\psi^a(u) &=& \left(1- (2 u -1)^2\right)\left(5 (2 u -1)^2 -1\right)
\frac{5}{2}
    \left(1 + \frac{9}{16} w^V_\gamma - \frac{3}{16} w^A_\gamma
    \right),
\nonumber \\
h_\gamma(u) &=& - 10 \left(1 + 2 \kappa^+\right) C_2^{\frac{1}{2}}(u
- \bar u),
\nonumber \\
\mathbb{A}(u) &=& 40 u^2 \bar u^2 \left(3 \kappa - \kappa^+
+1\right)  +
        8 (\zeta_2^+ - 3 \zeta_2) \left[u \bar u (2 + 13 u \bar u) \right.
\nonumber \\ && + \left.
                2 u^3 (10 -15 u + 6 u^2) \ln(u) + 2 \bar u^3 (10 - 15 \bar u + 6 \bar u^2)
        \ln(\bar u) \right],
\nonumber \\
{\cal A}(\alpha_i) &=& 360 \alpha_q \alpha_{\bar q} \alpha_g^2
        \left(1 + w^A_\gamma \frac{1}{2} (7 \alpha_g - 3)\right),
\nonumber \\
{\cal V}(\alpha_i) &=& 540 w^V_\gamma (\alpha_q - \alpha_{\bar q})
\alpha_q \alpha_{\bar q}
                \alpha_g^2,
\nonumber \\
{\cal T}_1(\alpha_i) &=& -120 (3 \zeta_2 + \zeta_2^+)(\alpha_{\bar
q} - \alpha_q)
        \alpha_{\bar q} \alpha_q \alpha_g,
\nonumber \\
{\cal T}_2(\alpha_i) &=& 30 \alpha_g^2 (\alpha_{\bar q} - \alpha_q)
    \left((\kappa - \kappa^+) + (\zeta_1 - \zeta_1^+)(1 - 2\alpha_g) +
    \zeta_2 (3 - 4 \alpha_g)\right),
\nonumber \\
{\cal T}_3(\alpha_i) &=& - 120 (3 \zeta_2 - \zeta_2^+)(\alpha_{\bar
q} -\alpha_q)
        \alpha_{\bar q} \alpha_q \alpha_g,
\nonumber 
\end{eqnarray}
\begin{eqnarray}
{\cal T}_4(\alpha_i) &=& 30 \alpha_g^2 (\alpha_{\bar q} - \alpha_q)
    \left((\kappa + \kappa^+) + (\zeta_1 + \zeta_1^+)(1 - 2\alpha_g) +
    \zeta_2 (3 - 4 \alpha_g)\right),\nonumber \\
{\cal S}(\alpha_i) &=& 30\alpha_g^2\{(\kappa +
\kappa^+)(1-\alpha_g)+(\zeta_1 + \zeta_1^+)(1 - \alpha_g)(1 -
2\alpha_g)\nonumber +\zeta_2[3 (\alpha_{\bar q} - \alpha_q)^2-\alpha_g(1 - \alpha_g)]\},\nonumber \\
\tilde {\cal S}(\alpha_i) &=&-30\alpha_g^2\{(\kappa -\kappa^+)(1-\alpha_g)+(\zeta_1 - \zeta_1^+)(1 - \alpha_g)(1 -
2\alpha_g)
+\zeta_2 [3 (\alpha_{\bar q} -\alpha_q)^2-\alpha_g(1 - \alpha_g)]\},
\end{eqnarray}
where $f_{3\gamma}=-0.0039~$GeV$^2$, $\zeta_2 = 0.3$,  $\varphi_2(1~GeV) = 0$, 
$w^V_\gamma = 3.8 \pm 1.8$, $w^A_\gamma = -2.1 \pm 1.0$, $\kappa = 0.2$, $\kappa^+ = 0$, $\zeta_1 = 0.4$,  and $\chi=-2.85 \pm 0.5~\mbox{GeV}^{-2}$~\cite{Rohrwild:2007yt}.
 
 \end{widetext}

 \bibliographystyle{elsarticle-num}
\bibliography{Zb-diquark.bib}
 
\end{document}